\documentstyle[11pt,newpasp]{article}

\begin{document}

\def\etal{{\sl et al.}}

\def\plotfiddle#1#2#3#4#5#6#7{\centering \leavevmode
\vbox to#2{\rule{0pt}{#2}}
\includegraphics{#1}}
%
%

\title{The Evolution of Spiral Disks}

\author{Matthew A. Bershady$^1$ and David R. Andersen$^2$}

\affil{$^1$Astronomy Department, University of Wisconsin, Madison, WI
53706 \\ $^2$Astron. \& Astrophys. Dept., Penn State, University Park,
PA, 16802}



\begin{abstract}
We report on aspects of an observational study to probe the mass
assembly of large galaxy disks. In this contribution we focus on a new
survey of integral-field H$\alpha$ velocity-maps of nearby, face on
disks. Preliminary results yield disk asymmetry amplitudes consistent
with estimates based on the scatter in the local Tully-Fisher
relation. We also show how the high quality of integral-field echelle
spectroscopy enables determinations of kinematic inclinations to
$i\sim20\deg$. This holds the promise that nearly-face-on galaxies can
be included in the Tully-Fisher relation. Finally, we discuss the
prospects for measuring dynamical asymmetries of distant galaxies.
\vskip -0.2in
\end{abstract}


\section{Introduction}

Tully-Fisher (TF) surveys at cosmological distances provide a direct
way to track the evolution in fundamental scaling relations of disk
galaxies. Such surveys permit, for example, the measurement of the
disk size or luminosity evolution under the assumption that deviations
from fiducial TF relations can be understood as photometric changes in
constant-mass potentials. Deviations from TF may also be produced by
dynamical asymmetries due to instabilities, interactions, minor
mergers, or slow and possibly lumpy mass-accretion.  Semi-analytic
models of galaxy formation are able to reproduce the basic slope and
scatter in the TF relationship at $z=0$, yet most fail to recover the
correct zero-point (e.g. Steinmetz \& Navarro, 1999). We anticipate
that in the future, however, the scatter in the TF relation will be a
more useful diagnostic of how disks are assembled. The scatter about
the TF relation should constrain the modes by which disks form,
particularly if the nature of this scatter can be determined as a
function of redshift. For example, are offsets from a fiducial TF
relation accompanied by an increase in disk asymmetry?

\section{Towards a Tully-Fisher Relation for Face-on Galaxies}

\begin{figure}
\plotfiddle{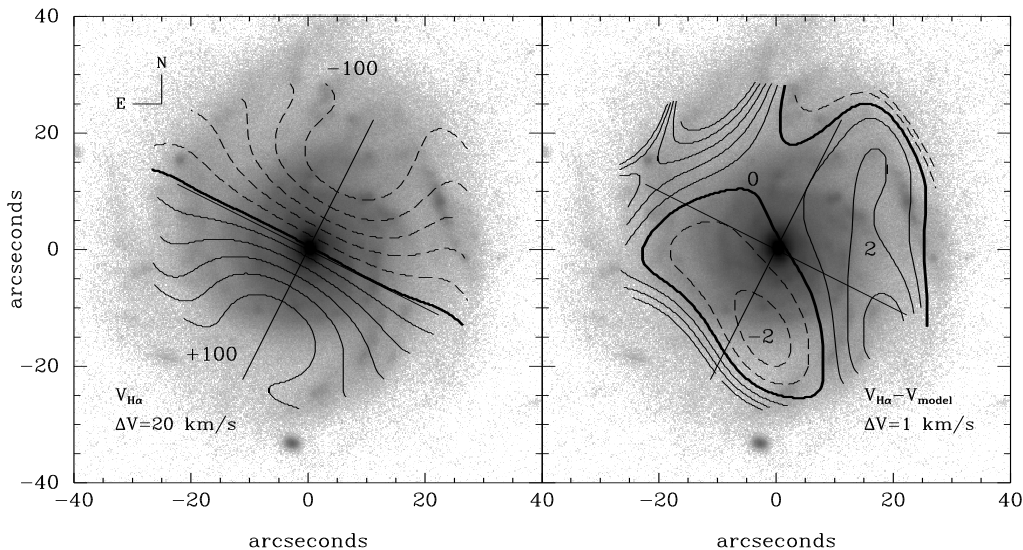}{2.25in}{-90}{60}{60}{-230}{280}
\vskip 0.15in
\caption{\hsize 5.25in WIYN/Densepak H$\alpha$ velocity-field for PGC 46767 ({\bf
left}), and the residuals from a single, inclined, rotating disk model
({\bf right}). The kinematic major and minor axes are drawn for
reference. Note the {\it low} amplitude, yet coherent structure of
the velocity residuals. Also note the large projected velocity for a
(photometrically) apparently face-on disk.}
\vskip -0.15in
\end{figure}

Historically, TF surveys have targeted galaxies with photometric
inclinations above 40-45$\deg$ in order to minimize errors in the
corrected rotation velocities. HI kinematic estimates of inclination
also have been difficult to measure below $i\sim40\deg$ given the
precision of HI maps in the past, as well as the intrinsic problems
associated with flaring and warping of the outer parts of HI disks
(e.g. Begeman, 1989). We have recently acquired integral-field echelle
spectroscopy with the WIYN telescope's Densepak fiber-bundle of over a
dozen, nearby disk galaxies ($0.02<z<0.05$). An example is shown in
Figure 1. In as little as an hour apiece we have obtained high
quality, H$\alpha$ velocity fields at a precision of $\sim2$ km/s over
the inner 2-3 few scale lengths of each disk. Moreover, we have been
successful fitting single, inclined disks to these kinematic data;
residuals are typically $\sim$3.5 km/s (rms). The derived kinematic
inclinations compare favorably to those estimated from inverting the
optical-HI TF relation (Figure 2, right panel). We estimate that the
kinematic inclinations from WIYN/Densepak H$\alpha$ maps can be
measured to better than 10\% down to $i=25\deg$; the dominant error
appears to arise from non-circular motions in the disks (Figure
3). The excellent correlation in Figure 2 (right panel) indicates,
however, that our error estimates may be substantially too
conservative!  The ability to study galaxies at low, but precisely
known inclinations is of interest for TF work because (i) detailed
disk structure can be viewed with little projection, and (ii) the
vertical component of the disk velocity ellipsoid is favorably
projected for measurement (potentially enabling estimates of both
total and disk mass).

\section{Disk Asymmetry in the Nearby Universe}

\begin{figure}
\plotfiddle{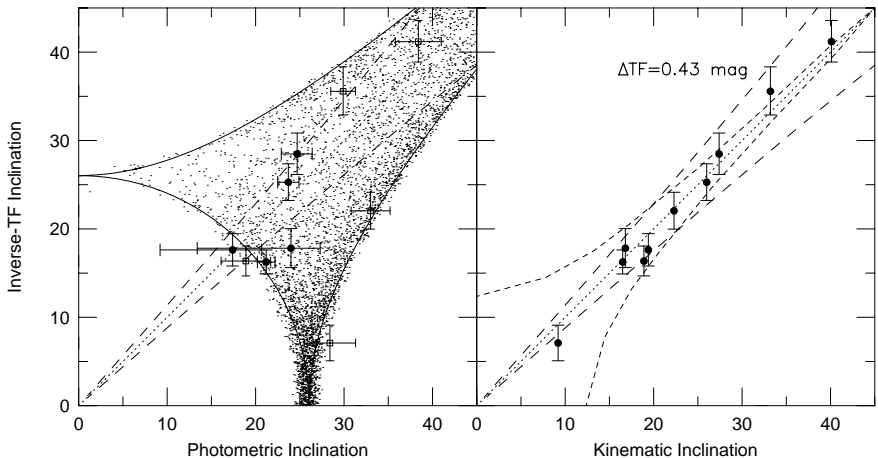}{1.85in}{0}{50}{50}{-180}{-75}
\vskip 0.15in
\caption{\hsize 5.25in A comparison of inclination angles derived from
inverting the TF relation (given observed HI line-width from Paturel
et al. 1997, and $B$ apparent magnitude from the RC3) versus: (i)
photometric inclinations based on WIYN $R$ and $I$ band images ({\bf
left} panel); and (ii) inclincations from modeling two-dimensional
H$\alpha$ velocity fields from WIYN/Densepak ({\bf right} panel).
Short and long-dashed lines are expected 1-$\sigma$ uncertainties for
kinematic and inverse-TF inclinations, respectively. Solid lines (left
panel) are the region accessible for disks with 10\% intrinsic
ellipticity; dots represent a Monte Carlo representation of a random
sampling of such disks.}
\vskip -0.15in
\end{figure}

By combining two-dimensional, kinematic information with optical and
near-infrared images of disks, one can explore in detail the asymmetry
of disks. The kinematic exploration has long been the domain of HI
studies (e.g. Richter \& Sanchisi 1994, Haynes et al. 1998, Swaters et
al. 1998), but now this can be done routinely in HII and folded more
directly into optical studies of photometric asymmetries
(e.g. Zaritsky \& Rix 1997; Kornreich et al. 1998; Conselice et
al. 2000). The next step, of course, is to be able to make comparable
kinematic measurements for the {\it stellar} component of disks.

Comparisons of the kinematic to photometric inclination and position
angles can yield constraints on the intrinsic ellipticity of disks. As
a simple test, we compare inclination angles in Figure 2. If disks are
intrinsically elliptic at the 10\% level (a limit inferred from the
scatter in the TF relation; Franx \& de Zeeuw, 1992), then there will
be an overabundance of apparently (photometrically) inclined galaxies
which are kinematically close to face-on. The small sample studied to
date is consistent with intrinsic disk ellipticities $\leq$10\%. More
powerful constraints lie in extracting apparent ellipticities relative
to kinematic major and minor axes (Andersen et al., in preparation).

\section{Measuring Disk Asymmetry at Higher Redshifts}

Velocity fields of distant disk galaxies would provide direct evidence
of their evolving dynamical states. However, the combined distance
effects of shrinking size, cosmological surface-brightness dimming,
and redshifting of spectral features into spectral regions of higher
terrestrial sky-backgrounds makes such observations daunting. Assuming
that one samples the same physical scale at all distances, the product
of the telescope diameter and observing-time$^{1/2}$ goes roughly as
$(1+z)^{4.25}$ from the ground, in the background limit, at high
spectral resolution, and at $z>0.7$ (where apparent size is a weak
function of redshift). Only modest gains ($-1.25$ in the exponent) are
made from space.

A zero-point for this telescope-diameter--time--redshift relation
can be set from the Keck observations of rotation curves at $z\sim1$,
taken in several hours with a 1 arcsec wide slit (Vogt et al. 1997). A
factor of 3 higher spatial resolution would just resolve galaxies at
cosmological distances at the equivalent of one disk scale-length for a
typical disk today. In the background limit, one would then require a
three times larger aperture at constant exposure time, i.e. a 30-m
telescope. Other gains can come from (i) improved instrument
throughput, and (ii) enhancing surface-brightness contrast by
resolving bright, star-forming knots in distant galaxy disks via
high-order adaptive optics (Koo, 1999). The latter is not applicable,
however, for the study of stellar motions in disks, which are of
particular interest for understanding asymmetries in disk mass
distributions.

\begin{figure}
\plotfiddle{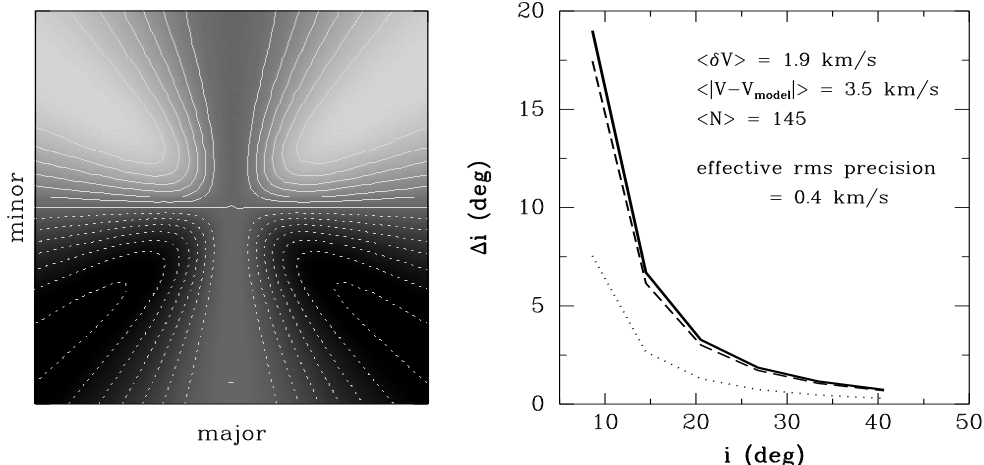}{1.85in}{-90}{60}{60}{-235}{270}
\vskip 0.1in
\caption{\hsize 5.25in The velocity-field residuals ({\bf left})
between two rotating disks at inclinations of 10$\deg$ and 20$\deg$
shows the maximum sensitivity to inclination lies at $\theta\sim\pm45$
from the kinematic major axis. For our characteristic instrumental
precision ($\delta$V), observed residuals (V-V$_{model}$), and number
of beams per galaxy (fibers, N), the {\bf right} panel shows our
estimated precision ($\Delta$i) as a function of true inclination for
deriving kinematic inclinations from WIYN/Densepak H$\alpha$ echelle
spectroscopy for our nearby galaxy sample.}
\vskip -0.15in
\end{figure}

\acknowledgments
We are grateful to acknowledge our collaborators in these projects --
L. S. Sparke, J. S. Gallagher, III, E. M. Wilcots, W. van Driel,
C.J. Conselice, and D. Ragaigne; M. Haynes, R. Giovanelli, C. Mihos,
and D. Koo. This research was supported by AST-96-18849 and
AST-99-70780.

\end{document}